\documentclass[11pt]{article}

\usepackage[margin=1in]{geometry}
\usepackage[T1]{fontenc}
\usepackage[utf8]{inputenc}
\usepackage{lmodern}
\usepackage{microtype}
\usepackage{amsmath,amssymb}
\usepackage{graphicx}
\usepackage{booktabs}
\usepackage{longtable}
\usepackage{array}
\usepackage{calc}
\usepackage{tabularx}
\usepackage{caption}
\usepackage{float}
\usepackage{url}
\usepackage{hyperref}
\usepackage{xcolor}
\usepackage{enumitem}
\usepackage{authblk}

\hypersetup{
  colorlinks=true,
  linkcolor=blue,
  citecolor=blue,
  urlcolor=blue,
  pdftitle={Persistent AI Agents in Academic Research: A Single-Investigator Implementation Case Study with Supplementary Appendix},
  pdfauthor={Anas H. Alzahrani, MD, PhD, MPH}
}

\setlist{nosep}
\providecommand{\tightlist}{\setlength{\itemsep}{0pt}\setlength{\parskip}{0pt}}
\makeatletter
\def\maxwidth{\ifdim\Gin@nat@width>\linewidth\linewidth\else\Gin@nat@width\fi}
\def\maxheight{\ifdim\Gin@nat@height>\textheight\textheight\else\Gin@nat@height\fi}
\makeatother
\setkeys{Gin}{width=\maxwidth,height=\maxheight,keepaspectratio}

\title{Persistent AI Agents in Academic Research: A Single-Investigator Implementation Case Study}
\author[1]{Anas H. Alzahrani, MD, PhD, MPH}
\affil[1]{Department of Preventive Medicine and Public Health, Faculty of Medicine, King Abdulaziz University, Jeddah, Saudi Arabia\\ORCID: \href{https://orcid.org/0000-0002-1394-9157}{0000-0002-1394-9157}\\\texttt{ahalzhrani@kau.edu.sa}}
\date{}

\begin{document}
\maketitle

\begin{abstract}

\textbf{Background:} Large language model systems are commonly evaluated as models, benchmarks, or short conversational episodes. Less is known about what happens when a persistent AI agent is embedded into a real academic research environment with durable memory, local files, external tools, scheduled routines, delegated roles, and explicit safety protocols.

\textbf{Objective:} To describe the implementation, utilization, outputs, resource profile, and governance layer of a persistent agentic research environment used by a single academic physician-scientist over 115 days.

\textbf{Methods:} This was a structured self-observed implementation case study conducted from January 31 to May 25, 2026. The unit of analysis was the human-agent environment: the researcher, agent runtime, memory files, tool access, repositories, scheduled jobs, specialized agent roles, and safety protocols. Data sources included recoverable session telemetry, memory files, repository and file-system inventories, model-use logs, decision logs, and documented protocol updates. Outcomes were descriptive and organized using PARE-M (Persistent Agentic Research Environment Measurement), a measurement framework covering system architecture, utilization, artifact production, resource use, reproducibility, and governance/correction domains.

\textbf{Results:} Recoverable main-agent telemetry included 75,671 de-duplicated records across 96 active days, including 8,059 user-role messages, 23,710 assistant-role messages, 18,596 tool-result messages, 2,385 tool-call events, and 1,286 model-completed events. The workspace inventory identified 502 memory-related files, 17 configured agent directories, and 57 skill files. Active system time was estimated at 579.7 hours using a 30-minute capped-gap rule and 674.1 hours using a 60-minute capped-gap rule. Memory-derived records identified 482 output-proxy events and 889 failure, verification, correction, or protocol-proxy events. A strict May 2026 trajectory subset captured 627 model-completed events and 73,950,305 recorded tokens, of which 61,278,669 were cache-read tokens (82.9\%); token telemetry was therefore interpreted as evidence of a cache-dominant persistent workflow rather than direct cash cost.

\textbf{Conclusions:} This case describes a persistent agentic research environment rather than an isolated chatbot interaction. The workflow was cache-dominant: 82.9\% of recorded May tokens were cache reads, suggesting that persistent agentic environments may shift the economic unit from cost per token to cost per completed artifact. Future evaluations should use artifact-level denominators, reproducible parsing rules, correction taxonomies, cost-per-artifact estimates, and independent coding of governance events.

\end{abstract}

\noindent\textbf{Keywords:} persistent AI agents; PARE-M; academic research; implementation case study; research workflow; token telemetry; agent governance

\section{Introduction}

Current evaluations of large language models emphasize model performance, benchmark scores, or task-level accuracy. This has been productive for comparing systems, but it leaves a gap: many real users no longer interact with AI as a single prompt-response episode. They embed agents into continuing work environments that include memory, files, tools, scripts, calendars, repositories, workflows, and institutional constraints.

This distinction matters in academic research. A physician-scientist does not only ask isolated questions. Research work involves manuscript development, study design, teaching, literature synthesis, code review, software deployment, collaboration, governance, and repeated correction. In this setting, the relevant object is not only ``the model.'' It is the coupled human-agent environment.

Prior work has described foundation models and their risks, tool-using and reasoning-acting agents, memory-augmented systems, and agent benchmarks \cite{ref1,ref2,ref3,ref4,ref5,ref6,ref7,ref8}. Recent work has also moved toward production-oriented agent memory, memory operating systems, memory-specific benchmarks, and multi-platform state-tracking evaluation \cite{ref12,ref13,ref14,ref15}. Implementation-science frameworks emphasize that technologies are shaped by users, organizations, context, and adaptation over time \cite{ref9,ref10,ref11}. To the author's knowledge, however, there are still few practice-based quantitative descriptions of persistent AI agents operating inside real academic research workflows.

This manuscript reports a 115-day structured self-observed implementation case study of such an environment. The goal is descriptive: to quantify what was built, how it was used, what outputs were recorded, what resources were consumed, and how safety/governance protocols accumulated. The study does not estimate causal productivity effects. It instead provides an empirical system description and introduces PARE-M (Persistent Agentic Research Environment Measurement), a measurement framework for future prospective evaluation.

\section{Case Context}

The setting was the working environment of a single academic physician-scientist in preventive medicine, public health, clinical research methods, and health data science. The workflow included manuscript drafting, causal-inference teaching, research-tool development, software deployment, literature review, administrative writing, product prototyping, and operational monitoring.

The interaction surface was a Discord-based channel connected to a local workspace. The environment included durable memory files, shell and file-system access, project repositories, scheduled jobs, external APIs, specialized agent roles, and rules governing external actions. The human researcher retained responsibility for scientific judgment, authorship, institutional accountability, and public-facing decisions.

Aqrab.ai and Coefficients Health Analytics are exploratory research-methodology projects associated with the author. Aqrab.ai appeared as one artifact domain in the workspace inventory and is therefore disclosed below as a contextual artifact category, not as independent evidence of scientific efficacy or as evidence of commercial performance.

\section{Methods}

\subsection{Design and Unit of Analysis}

This was a structured self-observed implementation case study over January 31--May 25, 2026. The unit of analysis was the persistent agentic research environment, defined as the combined human user, agent runtime, memory layer, tools, repositories, scheduled jobs, task protocols, and governance rules.

The study is not a controlled productivity trial. There was no randomized comparison, no matched non-agent workflow, and no prospective baseline. Comparative claims were therefore avoided unless they referred to explicitly observed within-system patterns.

\subsection{Data Sources and Operational Definitions}

Data sources are summarized in Table 1. ``Recoverable telemetry'' refers to locally available session and trajectory records after de-duplication. A ``de-duplicated record'' was a unique event retained after removing repeated records produced by overlapping logs. An ``active day'' was a calendar day with at least one recoverable main-agent event. A ``JSONL-like file'' was a line-oriented session or trajectory file parsable as event records, even when not every line conformed to strict JSONL. A ``skill file'' was a local instruction file defining a reusable procedure. A ``configured agent directory'' was a local role-specific agent folder available to the runtime. An ``output-proxy event'' was a dated memory entry indicating completion or delivery of a manuscript, teaching artifact, software change, deployment, protocol, analysis, or operational deliverable. The analytic observation window began on January 31, 2026, but recoverable telemetry starts on February 2, 2026; active-day counts and role-event counts are therefore lower-bound estimates for the full 115-day case.

\subsection{Measures}

Measures were organized using PARE-M v0.1, a descriptive measurement framework for persistent agentic research environments. PARE-M v0.1 separates utilization, output, resource, reproducibility, and governance domains and requires that each reported metric include a numerator, denominator, time window, and computation rule.

\paragraph{The PARE-M Framework (v0.1)}

\begin{longtable}{@{}
  >{\raggedright\arraybackslash}p{(\columnwidth - 6\tabcolsep) * \real{0.2308}}
  >{\raggedright\arraybackslash}p{(\columnwidth - 6\tabcolsep) * \real{0.2308}}
  >{\raggedleft\arraybackslash}p{(\columnwidth - 6\tabcolsep) * \real{0.3077}}
  >{\raggedright\arraybackslash}p{(\columnwidth - 6\tabcolsep) * \real{0.2308}}@{}}
\toprule\noalign{}
\begin{minipage}[b]{\linewidth}\raggedright
Metric
\end{minipage} & \begin{minipage}[b]{\linewidth}\raggedright
Definition
\end{minipage} & \begin{minipage}[b]{\linewidth}\raggedleft
Unit
\end{minipage} & \begin{minipage}[b]{\linewidth}\raggedright
Computation rule
\end{minipage} \\
\midrule\noalign{}
\endhead
\bottomrule\noalign{}
\endlastfoot
Active-day fraction (ADF) & Active days divided by calendar days in the observation window & Proportion & Count calendar days with at least one recoverable main-agent event and divide by total calendar days \\
De-duplicated record count (DRC) & Unique recoverable records after the Supplementary Methods S1.1 de-duplication rule & Count & Apply stable event/message identifiers first; otherwise hash timestamp, role, event type, content prefix, and tool name \\
Active-time estimate (ATE) & Capped-gap sum across unique event timestamps & Hours & Sort unique timestamps and sum consecutive gaps capped at 30 minutes; report 60-minute sensitivity \\
Cache-dominance ratio (CDR) & Cache-read tokens divided by total recorded tokens & Proportion & Sum cache-read tokens and divide by total input, output, cache-read, and cache-write tokens in the telemetry subset \\
Output-proxy rate (OPR) & Output-proxy events divided by active days & Events per active day & Count memory-recorded completed/delivered artifact events and divide by active-day count \\
Governance-event rate (GER) & Failure, verification, correction, or protocol-proxy events divided by active days & Events per active day & Count memory-recorded governance/protocol proxy events and divide by active-day count \\
Artifact-surface breadth (ASB) & Distinct artifact-surface categories represented in the workspace inventory & Count & Count unique artifact domains after mapping files to stable surface categories \\
\end{longtable}

\paragraph{Utilization}

Utilization measures included ADF, DRC, message-role counts, tool-call events, model-completed events, and ATE (Table 2). Active system time was estimated from unique event timestamps by summing gaps capped at 30 minutes, with a 60-minute cap used as sensitivity analysis.

\paragraph{Resources}

Resource measures included May 1--May 25, 2026 token telemetry, point-in-time VPS runtime characteristics, backup/recovery artifacts, and preliminary cost-accounting fields. Token categories were analyzed separately as input, output, cache-read, and cache-write tokens. Token telemetry was not treated as invoice cost, and VPS/API cost was reserved for invoice reconciliation rather than inferred from server specifications.

\paragraph{Outputs}

Output measures used two layers: memory-recorded output-proxy events and selected artifact-surface counts. File counts were interpreted cautiously because projects, manuscripts, static sites, and generated build products are not commensurable. Detailed inventories are therefore placed in the supplement rather than used as headline evidence.

\paragraph{Governance}

Governance measures included failure, verification, correction, and protocol-proxy events. These events were identified from memory and lesson files using keyword/rule-based extraction. They are author-coded and require independent validation before being treated as definitive rates.

\subsection{Reproducibility}

The reproducible components are the data schema, parsing rules, de-duplication logic, active-time algorithm, file-classification rules, and output-event extraction logic. Privacy constraints prevent release of raw conversations, credentials, unpublished collaborator material, and private project logs. A de-identified event ledger and parsing scripts are in preparation for preprint release (Supplementary Methods S1).

\section{Results}

\subsection{Architecture and Utilization}

By the end of follow-up, the environment included persistent memory, interaction channels, shell/file access, repositories, scheduled jobs, external APIs, specialized agent roles, and governance protocols (Table 1; Supplementary Figure S1). The inventory identified 502 memory-related files, 17 configured agent directories, 57 skill files, 4,309 main-session files, 3,194 recoverable main JSONL-like files, 5,760 all-agent session files, and 4,388 recoverable all-agent JSONL-like files.

Recoverable main-agent telemetry contained 75,671 de-duplicated records (DRC = 75,671) across 96 active days, corresponding to an active-day fraction of 0.835 (ADF = 96/115; Table 2). The longitudinal accumulation of activity, outputs, resources, and governance signals is summarized in Figure 2. Active system time was 579.7 hours using the primary 30-minute capped-gap estimate (ATE = 579.7 hours) and 674.1 hours using the 60-minute sensitivity estimate. These are system-activity estimates, not direct human labor hours.

\subsection{Outputs and Governance}

Memory-derived records identified 482 output-proxy events and 1,423 dated memory sections (Table 3). The output-proxy rate was 5.02 events per active day (OPR = 482/96). Selected artifact surfaces included manuscripts, teaching artifacts, content drafts, scripts, operational documents, software repositories, calibration-study materials, and deployed research tools; the supplementary inventory identified 10 artifact-surface categories (ASB = 10). Because these categories are heterogeneous, the main result is not the raw file total. The defensible finding is breadth: the environment was used across research, teaching, software, content, governance, and operations rather than only text generation.

The same memory layer recorded 889 failure, verification, correction, or protocol-proxy events, corresponding to a governance-event rate of 9.26 events per active day (GER = 889/96; Table 3). These included deployment safeguards, external-action checks, credential-handling rules, citation-verification rules, and lessons from duplicate or unsafe actions. The governance layer therefore became part of the operating environment rather than an after-the-fact policy appendix.

\subsection{Resource Profile}

The strict May 1--May 25 trajectory subset captured 627 model-completed events and 73,950,305 recorded tokens (Table 2; Figure 1). Cache-read tokens represented 61,278,669 of the total, yielding a cache-dominance ratio of 0.829 (CDR = 82.9\%), compared with 10,697,394 input tokens, 754,633 output tokens, and 1,219,609 cache-write tokens. Provider routes included OpenAI Codex-route usage, OpenAI-route usage, and Anthropic-route usage.

The operating environment ran on a DigitalOcean VPS and included local backup/recovery artifacts (Table 2; Supplementary Table S6). A verified compressed local backup of the OpenClaw environment was created during the observation period, with a recorded checksum, and additional local session/configuration backups were present. This should be interpreted as recoverability infrastructure, not as disaster-recovery maturity: no automated offsite backup schedule or restore drill was verified.

The cache-heavy profile is an empirical signal that persisted context was central to the workflow. It also shows why per-token accounting alone is a poor proxy for value. Preliminary direct-spend tracking identified approximately US\$1,961 in observed system-related spending, but invoice-level reconciliation remained incomplete (Table 2; Supplementary Table S6). Future work should estimate cost per completed artifact, cost per verified workflow, and cost per avoided correction rather than only aggregate token volume.

\subsection{Efficiency Hypothesis}

Aggregate interaction volume did not demonstrate reduced human input. As the environment accumulated memory, tools, and procedures, the scope of delegated work expanded. The strongest observed pattern is therefore capacity expansion, not proven labor substitution. The testable hypothesis for future work is artifact-level efficiency: comparable outputs may require fewer user prompts, fewer corrections, lower active system time, or lower marginal cost after reusable memory and protocols are established.

\section{Discussion}

The principal finding is that persistent agentic infrastructure expanded the capacity and scope of academic work. This was not shown by subjective usefulness alone. It was visible across PARE-M v0.1 utilization and governance metrics: ADF = 0.835, DRC = 75,671, OPR = 5.02 events per active day, GER = 9.26 events per active day, and ASB = 10. The environment became a durable work system spanning research, teaching, software, operations, and safety rules.

The likely mechanism was accumulated context plus reusable procedures. The workspace contained 502 memory-related files, 17 configured agent directories, and 57 skill files. The May token profile strengthens this interpretation: CDR was 82.9\%, suggesting that the workflow increasingly depended on reused context rather than isolated fresh inference. This aligns with memory-augmented and stateful-agent work, including recent production-memory and memory-benchmark studies, but differs from benchmark settings that evaluate agents on bounded tasks rather than lived deployment \cite{ref3,ref4,ref5,ref6,ref7,ref8,ref12,ref13,ref14,ref15}.

Governance was not separate from capability. In a persistent environment, the same memory infrastructure that preserves task context also preserves safety rules, prior mistakes, citation requirements, deployment checks, and external-action boundaries. That matters for academic medicine because unpublished projects, collaborator communications, credentials, and public-facing claims can coexist in the same workspace. Implementation frameworks such as CFIR, RE-AIM, and NASSS are useful here because they treat technologies as adaptive sociotechnical interventions rather than static tools \cite{ref9,ref10,ref11}.

This case also clarifies why episodic benchmarks miss part of the phenomenon. HELM, SWE-bench, AgentBench, GAIA, and AI-scientist-style evaluations provide important information about model and agent capabilities \cite{ref2,ref5,ref6,ref7,ref8}. They do not directly measure whether a persistent agentic environment improves a researcher's longitudinal workflow, governance, reproducibility, or cost per artifact. Those outcomes require system-level telemetry and prospective denominators.

The token-market implication is nontrivial. If persistent academic workflows become cache-dominant, then the binding constraint may shift from raw generation cost toward integration, governance, privacy, reproducibility, and provider-routing reliability. For such systems, ``cost per token'' is less informative than cost per artifact, cost per verified action, and cost per safe external workflow.

\section{Limitations}

This is a single-investigator self-observed case. The author was simultaneously user, system designer, data source, analyst, and beneficiary. There was no control group, no pre-specified output register, no baseline productivity period, and no independent coder for governance events. Artifact categories were assembled retrospectively and represent a positive inventory; abandoned, failed, or never-started work was not systematically counted. File counts are especially vulnerable to inflation from software projects and generated artifacts. Token telemetry was detailed only for May 1--May 25 and should not be conflated with the full 115-day window or with cash cost. These limitations make causal productivity claims inappropriate. They do not eliminate the value of the case as a structured system description.

\section{Conclusion}

Across 115 days, a persistent AI agent embedded in an academic physician-scientist's workspace accumulated measurable infrastructure, utilization, outputs, resource consumption, and governance rules. The most defensible conclusion is not that the system proved efficiency. It is that persistent agentic research environments can expand capacity while creating new requirements for artifact-level measurement, reproducibility, cost accounting, and safety governance.

\section{Main Tables and Figure}

\subsection{Table 1. System Architecture and Data Sources}

\begin{longtable}{@{}
  >{\raggedright\arraybackslash}p{(\columnwidth - 8\tabcolsep) * \real{0.1765}}
  >{\raggedleft\arraybackslash}p{(\columnwidth - 8\tabcolsep) * \real{0.2353}}
  >{\raggedleft\arraybackslash}p{(\columnwidth - 8\tabcolsep) * \real{0.2353}}
  >{\raggedright\arraybackslash}p{(\columnwidth - 8\tabcolsep) * \real{0.1765}}
  >{\raggedright\arraybackslash}p{(\columnwidth - 8\tabcolsep) * \real{0.1765}}@{}}
\toprule\noalign{}
\begin{minipage}[b]{\linewidth}\raggedright
Domain
\end{minipage} & \begin{minipage}[b]{\linewidth}\raggedleft
Component or source
\end{minipage} & \begin{minipage}[b]{\linewidth}\raggedleft
Observed value
\end{minipage} & \begin{minipage}[b]{\linewidth}\raggedright
Analytic role
\end{minipage} & \begin{minipage}[b]{\linewidth}\raggedright
Caveat
\end{minipage} \\
\midrule\noalign{}
\endhead
\bottomrule\noalign{}
\endlastfoot
Observation window & Calendar interval & Jan 31--May 25, 2026; 115 days & Bounds the case & Recoverable telemetry starts Feb 2 \\
Memory & Memory-related files & 502 & Persistent context and governance & Includes heterogeneous memory artifacts \\
Agents & Configured agent directories & 17 & Role specialization & Does not imply equal activity \\
Skills & Skill files & 57 & Reusable procedures & File count, not quality measure \\
Sessions & Main-session files & 4,309 & Interaction record source & Completeness varies \\
Sessions & Recoverable main JSONL-like files & 3,194 & Parsed telemetry source & ``JSONL-like'' defined operationally \\
Sessions & Recoverable all-agent JSONL-like files & 4,388 & Multi-agent telemetry source & Main analysis uses main-agent records \\
Governance & Lessons/protocol files & Present & Safety and correction layer & Requires taxonomy validation \\
\end{longtable}

\subsection{Table 2. Utilization and Resource Metrics}

\begin{longtable}{@{}
  >{\raggedright\arraybackslash}p{(\columnwidth - 8\tabcolsep) * \real{0.1875}}
  >{\raggedleft\arraybackslash}p{(\columnwidth - 8\tabcolsep) * \real{0.2500}}
  >{\raggedright\arraybackslash}p{(\columnwidth - 8\tabcolsep) * \real{0.1875}}
  >{\raggedright\arraybackslash}p{(\columnwidth - 8\tabcolsep) * \real{0.1875}}
  >{\raggedright\arraybackslash}p{(\columnwidth - 8\tabcolsep) * \real{0.1875}}@{}}
\toprule\noalign{}
\begin{minipage}[b]{\linewidth}\raggedright
Metric
\end{minipage} & \begin{minipage}[b]{\linewidth}\raggedleft
Value
\end{minipage} & \begin{minipage}[b]{\linewidth}\raggedright
Window
\end{minipage} & \begin{minipage}[b]{\linewidth}\raggedright
Interpretation
\end{minipage} & \begin{minipage}[b]{\linewidth}\raggedright
Caveat
\end{minipage} \\
\midrule\noalign{}
\endhead
\bottomrule\noalign{}
\endlastfoot
Active days & 96 & Feb 2--May 25 & Sustained use & Lower-bound recoverable logs \\
De-duplicated records & 75,671 & Feb 2--May 25 & Overall telemetry volume & Depends on parser \\
User-role messages & 8,059 & Feb 2--May 25 & Human input events & Not equal to effort \\
Assistant-role messages & 23,710 & Feb 2--May 25 & Agent output/events & Includes operational messages \\
Tool-result messages & 18,596 & Feb 2--May 25 & Tool-mediated work & Not all are successes \\
Tool-call events & 2,385 & Feb 2--May 25 & External/local action use & Needs action taxonomy \\
Model-completed events & 1,286 & Feb 2--May 25 & Model invocations in broad recoverable main telemetry & Not identical to billed calls \\
Active system time & 579.7 hours & 30-min cap & Primary activity estimate & Not human labor time \\
Active system time & 674.1 hours & 60-min cap & Sensitivity estimate & Cap choice requires histogram \\
Strict model-completed trajectory events & 627 & May 1--May 25 & Reproducible token subset used for Figure 1 & Excludes broader session-level usage records \\
Recorded tokens & 73,950,305 & May 1--May 25 & Resource telemetry in strict trajectory subset & Partial month only \\
Cache-read tokens & 61,278,669; 82.9\% & May 1--May 25 & Cache-dominant workflow & Not invoice cost \\
Output tokens & 754,633 & May 1--May 25 & Generated token volume & Not artifact count \\
VPS host class & DigitalOcean Droplet, lon1; Ubuntu 24.04.4 LTS & Observed May 26, 2026 & Runtime infrastructure context & Invoice price not recovered locally \\
Runtime capacity & 4 vCPU; 7.8 GiB RAM; 154 GiB disk, 99 GiB used & Observed May 26, 2026 & Compute/storage burden & Point-in-time server state \\
Backup/recovery artifacts & 5.4 GB verified local compressed backup plus local session/config backups & Observed May 2026 & Recoverability infrastructure & No verified offsite backup or restore drill \\
Preliminary observed direct spend & Approximately US\$1,961 & Jan 31--May 25 & Cash-burden anchor for system operation & Requires invoice-level reconciliation \\
\end{longtable}

\subsection{Table 3. Output and Governance Ledger}

\begin{longtable}{@{}
  >{\raggedright\arraybackslash}p{(\columnwidth - 8\tabcolsep) * \real{0.1875}}
  >{\raggedleft\arraybackslash}p{(\columnwidth - 8\tabcolsep) * \real{0.2500}}
  >{\raggedright\arraybackslash}p{(\columnwidth - 8\tabcolsep) * \real{0.1875}}
  >{\raggedright\arraybackslash}p{(\columnwidth - 8\tabcolsep) * \real{0.1875}}
  >{\raggedright\arraybackslash}p{(\columnwidth - 8\tabcolsep) * \real{0.1875}}@{}}
\toprule\noalign{}
\begin{minipage}[b]{\linewidth}\raggedright
Domain
\end{minipage} & \begin{minipage}[b]{\linewidth}\raggedleft
Observed value
\end{minipage} & \begin{minipage}[b]{\linewidth}\raggedright
Interpretation
\end{minipage} & \begin{minipage}[b]{\linewidth}\raggedright
Main caveat
\end{minipage} & \begin{minipage}[b]{\linewidth}\raggedright
Supplement
\end{minipage} \\
\midrule\noalign{}
\endhead
\bottomrule\noalign{}
\endlastfoot
Dated memory sections & 1,423 & Structured longitudinal record & Not all are outputs & S1 \\
Output-proxy events & 482 & Completed or delivered work signals & Keyword/rule-derived & S2 \\
Manuscript surface & 39 files & Manuscript-related work & File count not artifact count & S3 \\
Teaching surface & 1,488 files & Teaching-material production & May include generated assets & S3 \\
Content surface & 54 files & LinkedIn/content drafts & Not all published & S3 \\
Script surface & 37 files & Reusable automation & File count not reuse count & S3 \\
Operations documents & 41 files & Workflow/governance support & Heterogeneous & S3 \\
Aqrab website source & 119 files & Research-methodology project surface & Context-relevant & S3 \\
Calibration/panel app surfaces & 61 files & Study infrastructure & Software files not study outcomes & S3 \\
Failure/protocol/verification proxies & 889 & Governance accumulation & Needs independent coding & S4 \\
\end{longtable}

\subsection{Figure 1. Token Telemetry and Cache-Output Relationship}

\begin{figure}
\centering
\includegraphics{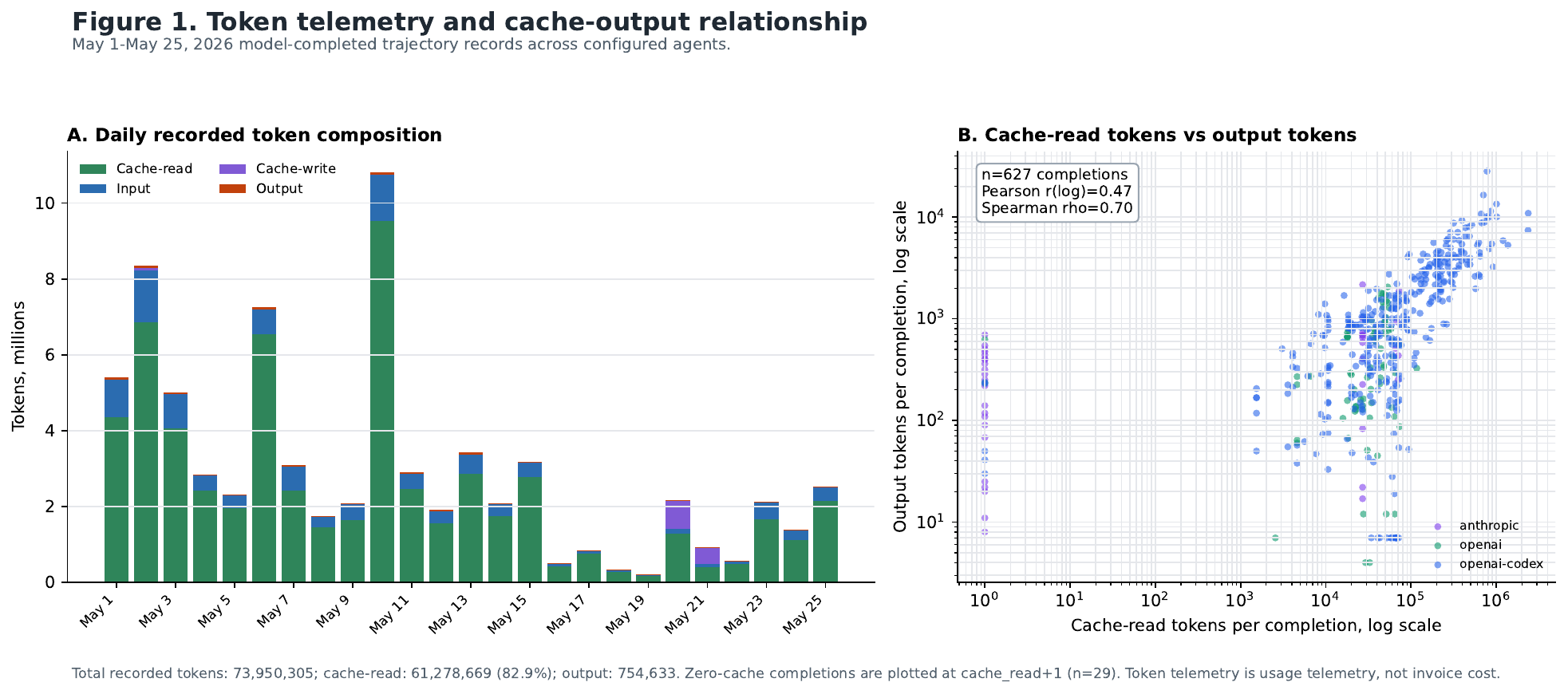}
\caption{Figure 1. Token telemetry and cache-output relationship}
\end{figure}

Panel A shows daily token composition from strict model-completed trajectory records. Panel B plots cache-read tokens against output tokens per completion on log scales. The observed association was positive (Pearson r on log-transformed counts = 0.47; Spearman rho = 0.70), supporting the interpretation that large generated outputs tended to occur in high-context, cache-heavy completions rather than isolated short exchanges.

Source files: \texttt{figures/figure-1-token-telemetry-cache-output.svg}, \texttt{figures/figure-1-token-telemetry-cache-output.pdf}, \texttt{figures/figure-1-token-telemetry-cache-output.pdf}, \texttt{figures/figure-1-token-telemetry-daily.csv}, and \texttt{figures/figure-1-token-telemetry-events.csv}.

\subsection{Figure 2. Longitudinal Evolution of the Environment}

\begin{figure}
\centering
\includegraphics{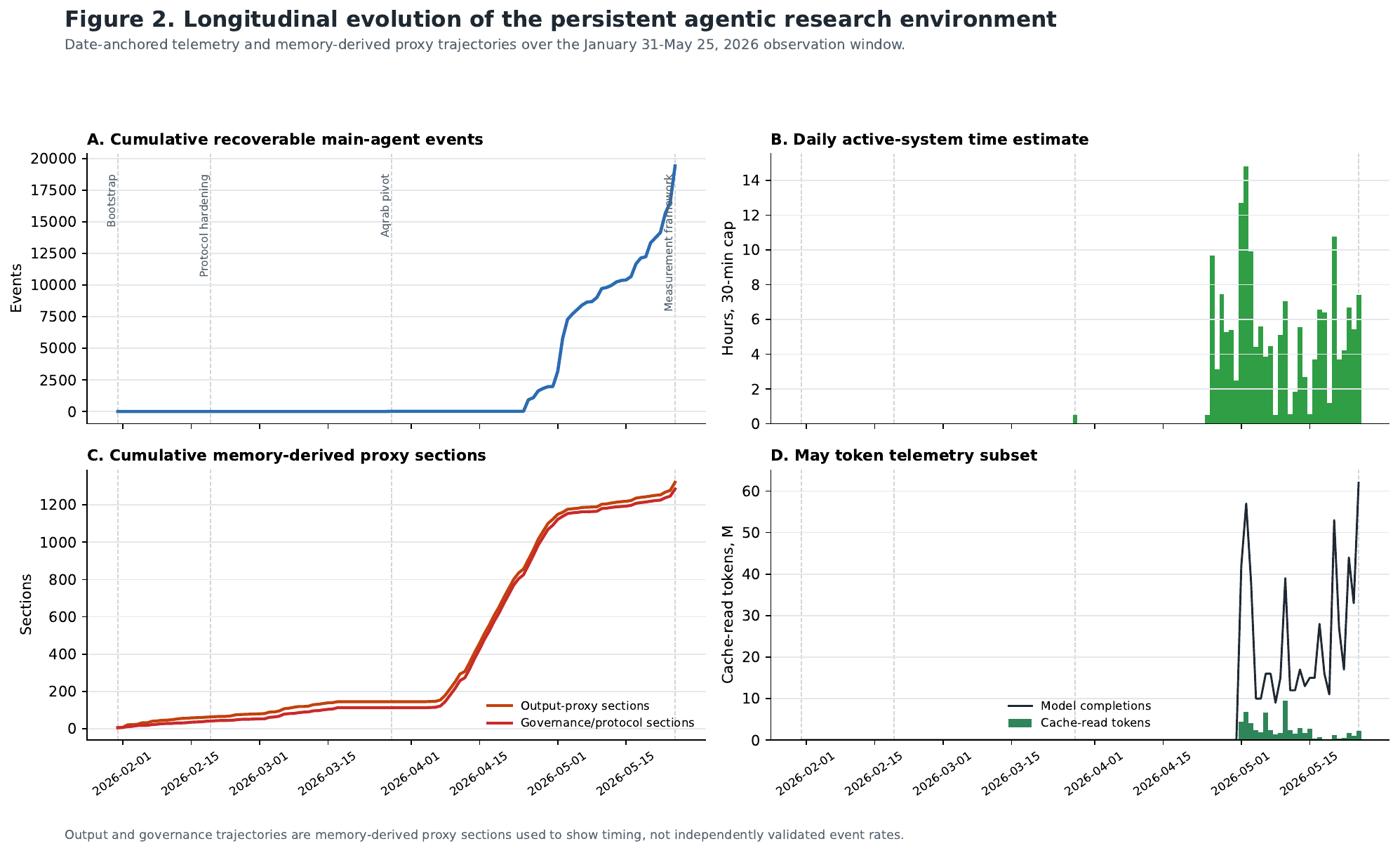}
\caption{Figure 2. Longitudinal evolution of the persistent agentic research environment}
\end{figure}

Figure 2 summarizes date-anchored longitudinal evolution of the environment across the observation window, including cumulative recoverable main-agent events, daily active-system time, memory-derived output/governance proxy sections, and the May token-telemetry subset.

Source files: \texttt{figures/figure-2-longitudinal-evolution.svg}, \texttt{figures/figure-2-longitudinal-evolution.pdf}, and \texttt{figures/figure-2-longitudinal-evolution.pdf}.

\subsection{Supplementary Conceptual Figures}

The conceptual architecture diagram has been moved to the supplement because it describes the mental model of the system rather than an observed outcome.

\section{Supplementary Materials}

Supplementary Methods S1: De-duplication key, active-time capped-gap pseudocode, file-classification rules, output-proxy extraction rules, and privacy/de-identification plan.\\
Supplementary Table S1: Full session and memory inventory.\\
Supplementary Table S2: Output-proxy event extraction categories and examples.\\
Supplementary Table S3: Artifact inventory mapped to logical artifacts rather than only files.\\
Supplementary Table S4: Failure, verification, correction, and protocol taxonomy with proposed second-coder validation.\\
Supplementary Table S5: Token/provider telemetry, provider-route cache dominance, and invoice-reconciliation plan.\\
Supplementary Table S6: Runtime infrastructure, backup/recovery artifacts, and cost-reconciliation fields.\\
Supplementary Figure S1: Persistent agentic research environment architecture.\\
Supplementary Figure S2: Distribution of event gaps and active-time sensitivity using 15-, 30-, 45-, 60-, and 90-minute caps.

\section{Reflexivity and Competing Interests}

This is a self-observed case study. The author created, used, corrected, and analyzed the environment. This creates unavoidable reflexivity and risk of favorable interpretation. Mitigation steps for the next version include publishing de-identified schemas and extraction rules, adding an independent audit of a subsample of governance events, and separating positive artifact inventories from attempted or abandoned work.

The author is associated with Aqrab.ai and Coefficients Health Analytics, exploratory research-methodology projects related to AI-assisted research methodology and evidence evaluation. Aqrab-related files were counted as an artifact surface in this case and are disclosed for transparency; they should not be interpreted as evidence of commercial performance.

\clearpage

\appendix

\section{Supplementary Appendix}

Manuscript: \emph{Persistent AI Agents in Academic Research: A Single-Investigator Implementation Case Study}

\section{Supplementary Methods S1. Reproducibility Schema}

\subsection{S1.1 De-duplication Rule}

Recoverable logs may contain overlapping records because the runtime stores session, trajectory, and message records in multiple locations. The de-duplicated analytic record should be generated using the first available stable key in this order:

\begin{enumerate}
\def\labelenumi{\arabic{enumi}.}
\tightlist
\item
  Explicit event/message identifier, if present.
\item
  Hash of \texttt{\{timestamp,\ role,\ event\_type,\ content\_prefix,\ tool\_name\}}.
\item
  Hash of \texttt{\{trajectory\_ts,\ provider\_route,\ model,\ token\_counts\}} for model-completed records.
\end{enumerate}

Records without timestamps should be excluded from active-time analysis but retained in file/source inventory counts if their source file is recoverable.

\subsection{S1.2 Active-Time Algorithm}

Primary estimate:

\begin{verbatim}
Input: sorted unique event timestamps for recoverable main-agent records
For each consecutive pair:
  gap = timestamp[i+1] - timestamp[i]
  contribution = min(gap, 30 minutes)
Sum contributions across all pairs
Report hours = sum / 3600
\end{verbatim}

Sensitivity estimates replace the cap with 15, 45, 60, and 90 minutes. Supplementary Figure S2 reports the raw inter-event gap distribution, active-system hours, and cluster counts under these caps.

\subsection{S1.3 Output-Proxy Extraction Rules}

An output-proxy event is a dated memory entry that documents completion, delivery, deployment, submission, publication, revision, analysis, verification, or creation of a substantive artifact. Candidate keyword families:

\begin{itemize}
\tightlist
\item
  \texttt{created}, \texttt{drafted}, \texttt{wrote}, \texttt{generated}, \texttt{rendered}
\item
  \texttt{implemented}, \texttt{fixed}, \texttt{patched}, \texttt{deployed}, \texttt{pushed}, \texttt{merged}
\item
  \texttt{verified}, \texttt{validated}, \texttt{smoke\ test}, \texttt{build\ passed}
\item
  \texttt{submitted}, \texttt{published}, \texttt{sent}, \texttt{posted}
\item
  \texttt{artifact}, \texttt{manuscript}, \texttt{slides}, \texttt{guide}, \texttt{script}, \texttt{dashboard}, \texttt{app}
\end{itemize}

Exclusions:

\begin{itemize}
\tightlist
\item
  Pure discussion without a delivered artifact.
\item
  Repeated logs of the same artifact unless a materially new version was produced.
\item
  Auto-generated build artifacts unless the analytic unit is explicitly a software surface.
\end{itemize}

\subsection{S1.4 File-Classification Rules}

File counts are artifact-surface indicators, not output counts. Classify files by stable workspace root and intended project role:

\begin{itemize}
\tightlist
\item
  \texttt{manuscripts/}: manuscript drafts, reports, responses.
\item
  \texttt{teaching-artifacts/}: teaching guides, visuals, slides, video scripts.
\item
  \texttt{linkedin/} and content folders: content drafts and social artifacts.
\item
  \texttt{scripts/}: reusable local automation.
\item
  \texttt{ops/}: operational documents and ledgers.
\item
  \texttt{aqrab-website/src/}: Aqrab product source surface.
\item
  \texttt{aqrab-calibration-study/research/}: calibration-study research surface.
\item
  \texttt{aqrab-calibration-study/panel-app/}: calibration panel app source surface.
\item
  \texttt{target-trial-emulation-benchmark/}: benchmark manuscript/data/software surface.
\end{itemize}

The next version should deduplicate these into logical artifacts, for example: ``one guide,'' ``one deployed app,'' ``one manuscript version,'' or ``one analysis pipeline.''

\section{Supplementary Table S1. Source Inventory}

\begin{longtable}{@{}
  >{\raggedright\arraybackslash}p{(\columnwidth - 6\tabcolsep) * \real{0.2308}}
  >{\raggedleft\arraybackslash}p{(\columnwidth - 6\tabcolsep) * \real{0.3077}}
  >{\raggedright\arraybackslash}p{(\columnwidth - 6\tabcolsep) * \real{0.2308}}
  >{\raggedright\arraybackslash}p{(\columnwidth - 6\tabcolsep) * \real{0.2308}}@{}}
\toprule\noalign{}
\begin{minipage}[b]{\linewidth}\raggedright
Source class
\end{minipage} & \begin{minipage}[b]{\linewidth}\raggedleft
Count or status
\end{minipage} & \begin{minipage}[b]{\linewidth}\raggedright
Used for
\end{minipage} & \begin{minipage}[b]{\linewidth}\raggedright
Limitation
\end{minipage} \\
\midrule\noalign{}
\endhead
\bottomrule\noalign{}
\endlastfoot
Daily memory files & 99 & Longitudinal chronology & Human/agent-authored notes, not raw telemetry \\
Memory-related files & 502 & Persistent context inventory & Heterogeneous \\
Main-session files & 4,309 & Main interaction source & Completeness varies \\
Recoverable main JSONL-like files & 3,194 & Parsed main telemetry & Operationally parsed, not guaranteed strict JSONL \\
All-agent session files & 5,760 & Multi-agent source frame & Not primary analytic set \\
Recoverable all-agent JSONL-like files & 4,388 & Multi-agent parsed frame & Includes agents outside main case narrative \\
Configured agent directories & 17 & Role specialization & Activity level not equal across agents \\
Skill files & 57 & Reusable procedure count & Count does not measure execution frequency \\
\end{longtable}

\section{Supplementary Table S2. Output-Proxy Categories}

\begin{longtable}{@{}
  >{\raggedright\arraybackslash}p{(\columnwidth - 4\tabcolsep) * \real{0.3333}}
  >{\raggedright\arraybackslash}p{(\columnwidth - 4\tabcolsep) * \real{0.3333}}
  >{\raggedright\arraybackslash}p{(\columnwidth - 4\tabcolsep) * \real{0.3333}}@{}}
\toprule\noalign{}
\begin{minipage}[b]{\linewidth}\raggedright
Category
\end{minipage} & \begin{minipage}[b]{\linewidth}\raggedright
Examples
\end{minipage} & \begin{minipage}[b]{\linewidth}\raggedright
Current status
\end{minipage} \\
\midrule\noalign{}
\endhead
\bottomrule\noalign{}
\endlastfoot
Manuscript/research writing & Drafts, revisions, reports, response letters & Counted as memory/output proxies and artifact files \\
Teaching artifacts & Guides, visuals, slides, videos, decision trees & Counted as artifact surface; logical dedup pending \\
Software/product work & Web apps, panel app, deployed tools, scripts & Counted as source surfaces; generated files should be excluded in final audit \\
Operations/governance & Policies, lessons, decision journals, safety rules & Counted as governance-relevant artifacts \\
Communication/content & LinkedIn drafts, outreach drafts, public posts & Counted only when recoverable in memory/files \\
\end{longtable}

\section{Supplementary Table S3. Artifact-Surface Counts}

\begin{longtable}{@{}
  >{\raggedright\arraybackslash}p{(\columnwidth - 8\tabcolsep) * \real{0.1765}}
  >{\raggedleft\arraybackslash}p{(\columnwidth - 8\tabcolsep) * \real{0.2353}}
  >{\raggedleft\arraybackslash}p{(\columnwidth - 8\tabcolsep) * \real{0.2353}}
  >{\raggedright\arraybackslash}p{(\columnwidth - 8\tabcolsep) * \real{0.1765}}
  >{\raggedright\arraybackslash}p{(\columnwidth - 8\tabcolsep) * \real{0.1765}}@{}}
\toprule\noalign{}
\begin{minipage}[b]{\linewidth}\raggedright
Surface
\end{minipage} & \begin{minipage}[b]{\linewidth}\raggedleft
Raw file count
\end{minipage} & \begin{minipage}[b]{\linewidth}\raggedleft
Logical artifact count after deduplication
\end{minipage} & \begin{minipage}[b]{\linewidth}\raggedright
Interpretation
\end{minipage} & \begin{minipage}[b]{\linewidth}\raggedright
Required cleanup
\end{minipage} \\
\midrule\noalign{}
\endhead
\bottomrule\noalign{}
\endlastfoot
Manuscripts & 39 files & Pending audit & Manuscript surface & Group into logical manuscripts/versions \\
Teaching artifacts & 1,488 files & Pending audit & Teaching surface & Exclude generated assets/build products where appropriate \\
LinkedIn/content drafts & 54 files & Pending audit & Content surface & Distinguish draft vs published \\
Revenue tools & 14 files & Pending audit & Revenue-validation surface & Link to lead/revenue outcomes \\
Scripts & 37 files & Pending audit & Reusable automation & Count actual reuse frequency \\
Operations documents & 41 files & Pending audit & Governance/ops surface & Separate policy from logs \\
Aqrab website source & 119 files & Pending audit & Exploratory research-methodology project source surface & Disclose context; do not treat as scientific output or commercial performance \\
Aqrab calibration research & 22 files & Pending audit & Calibration-study research surface & Link to protocol/OSF status \\
Panel app & 39 files & Pending audit & Study infrastructure & Software surface, not outcome \\
Target-trial benchmark & 80 files & Pending audit & Benchmark manuscript/data surface & Separate manuscript, data, and generated outputs \\
\end{longtable}

\section{Supplementary Table S4. Governance Event Taxonomy}

\begin{longtable}{@{}
  >{\raggedright\arraybackslash}p{(\columnwidth - 6\tabcolsep) * \real{0.2500}}
  >{\raggedright\arraybackslash}p{(\columnwidth - 6\tabcolsep) * \real{0.2500}}
  >{\raggedright\arraybackslash}p{(\columnwidth - 6\tabcolsep) * \real{0.2500}}
  >{\raggedright\arraybackslash}p{(\columnwidth - 6\tabcolsep) * \real{0.2500}}@{}}
\toprule\noalign{}
\begin{minipage}[b]{\linewidth}\raggedright
Proposed class
\end{minipage} & \begin{minipage}[b]{\linewidth}\raggedright
Definition
\end{minipage} & \begin{minipage}[b]{\linewidth}\raggedright
Examples
\end{minipage} & \begin{minipage}[b]{\linewidth}\raggedright
Validation need
\end{minipage} \\
\midrule\noalign{}
\endhead
\bottomrule\noalign{}
\endlastfoot
Verification event & Check performed before/after action & Build tests, DOI checks, deployment smoke tests & Extract exact denominator \\
Correction event & Error fixed after detection & Wrong default score, auth callback bug, citation correction & Code second sample \\
Protocol event & Durable rule added & External-action checklist, Vercel deploy rule, credential rule & Distinguish new vs repeated rules \\
Safety event & Potential harmful disclosure/action prevented or corrected & Credential handling, public-surface limits & De-identify before release \\
Failure event & Completed action with wrong/undesired result & Duplicate sends, bad DOI, broken deploy & Severity grading needed \\
\end{longtable}

The current aggregate count of 889 combines these classes and should not be used as a rate until classified.

\section{Supplementary Table S5. Token and Provider Telemetry}

\begin{longtable}{@{}
  >{\raggedright\arraybackslash}p{(\columnwidth - 6\tabcolsep) * \real{0.2308}}
  >{\raggedleft\arraybackslash}p{(\columnwidth - 6\tabcolsep) * \real{0.3077}}
  >{\raggedright\arraybackslash}p{(\columnwidth - 6\tabcolsep) * \real{0.2308}}
  >{\raggedright\arraybackslash}p{(\columnwidth - 6\tabcolsep) * \real{0.2308}}@{}}
\toprule\noalign{}
\begin{minipage}[b]{\linewidth}\raggedright
Metric
\end{minipage} & \begin{minipage}[b]{\linewidth}\raggedleft
Value
\end{minipage} & \begin{minipage}[b]{\linewidth}\raggedright
Window
\end{minipage} & \begin{minipage}[b]{\linewidth}\raggedright
Interpretation
\end{minipage} \\
\midrule\noalign{}
\endhead
\bottomrule\noalign{}
\endlastfoot
Strict model-completed trajectory events & 627 & May 1--May 25, 2026 & Reproducible subset used for Figure 1 \\
Total recorded tokens & 73,950,305 & May 1--May 25, 2026 & Usage telemetry, not invoice cost \\
Input tokens & 10,697,394 & May 1--May 25, 2026 & Fresh/context input \\
Output tokens & 754,633 & May 1--May 25, 2026 & Generated output \\
Cache-read tokens & 61,278,669 & May 1--May 25, 2026 & 82.9\% of recorded tokens \\
Cache-write tokens & 1,219,609 & May 1--May 25, 2026 & Context written to cache \\
OpenAI Codex route & 68.8M tokens & May 1--May 25, 2026 & Provider route, not necessarily invoice line \\
OpenAI Codex route CDR & 84.1\% & May 1--May 25, 2026 & Cache-read share within this route \\
OpenAI route & 3.2M tokens & May 1--May 25, 2026 & Provider route \\
OpenAI route CDR & 84.6\% & May 1--May 25, 2026 & Cache-read share within this route \\
Anthropic route & 1.9M tokens & May 1--May 25, 2026 & Provider route \\
Anthropic route CDR & 35.2\% & May 1--May 25, 2026 & Cache-read share within this route; lower because cache-write tokens were concentrated here \\
\end{longtable}

The strict subset is generated from \texttt{model.completed} events in local OpenClaw trajectory files and exported as \texttt{figures/figure-1-token-telemetry-daily.csv} and \texttt{figures/figure-1-token-telemetry-events.csv}.

\section{Supplementary Table S6. Runtime Infrastructure, Backup, and Cost-Reconciliation Fields}

\begin{longtable}{@{}
  >{\raggedright\arraybackslash}p{(\columnwidth - 6\tabcolsep) * \real{0.2308}}
  >{\raggedleft\arraybackslash}p{(\columnwidth - 6\tabcolsep) * \real{0.3077}}
  >{\raggedright\arraybackslash}p{(\columnwidth - 6\tabcolsep) * \real{0.2308}}
  >{\raggedright\arraybackslash}p{(\columnwidth - 6\tabcolsep) * \real{0.2308}}@{}}
\toprule\noalign{}
\begin{minipage}[b]{\linewidth}\raggedright
Field
\end{minipage} & \begin{minipage}[b]{\linewidth}\raggedleft
Observed value
\end{minipage} & \begin{minipage}[b]{\linewidth}\raggedright
Source/Timing
\end{minipage} & \begin{minipage}[b]{\linewidth}\raggedright
Interpretation
\end{minipage} \\
\midrule\noalign{}
\endhead
\bottomrule\noalign{}
\endlastfoot
Host class & DigitalOcean Droplet, lon1 & Local runtime context and hostname/inventory notes & VPS infrastructure used for the observed environment \\
Operating system & Ubuntu 24.04.4 LTS & \texttt{lsb\_release}, May 26, 2026 & Runtime operating system \\
Kernel & Linux 6.8.0-111-generic x86\_64 & \texttt{uname}, May 26, 2026 & Runtime kernel \\
CPU capacity & 4 vCPU observed & \texttt{nproc}, May 26, 2026 & Compute capacity \\
Memory capacity & 7.8 GiB total RAM & \texttt{free\ -h}, May 26, 2026 & Memory burden \\
Disk capacity & 154 GiB root disk; 99 GiB used & \texttt{df\ -h\ /}, May 26, 2026 & Storage burden \\
OpenClaw installation/workspace footprint & 16 GB & \texttt{du\ -sh}, May 26, 2026 & Approximate local environment size \\
Verified local backup & 5.4 GB compressed OpenClaw backup archive & Local backup archive and checksum sidecar, May 5, 2026 & Point-in-time compressed recovery artifact \\
Backup checksum & SHA256 sidecar present for the May 5 archive & Local backup directory, May 5, 2026 & Integrity check support \\
Session/config backups & Session backup folders and \texttt{openclaw.json} backup/last-good files present & Local OpenClaw directory, May 2026 & Partial operational recovery layer \\
Offsite backup & Not verified & Local inspection only & Disaster-recovery limitation \\
Restore drill & Not verified & Local inspection only & Recoverability not proven end-to-end \\
Preliminary observed direct spend & Approximately US\$1,961 & Memory/cost-tracking summary through May 25, 2026 & Direct-spend anchor; not yet invoice-audited \\
VPS invoice cost & Not recovered locally & Requires DigitalOcean invoice & Should be reported as cash cost only after invoice reconciliation \\
OpenRouter account usage & 134.49 credits used from 140 total credits & Provider account check, May 2026 & Provider-credit burden, not necessarily invoice cash \\
ElevenLabs account & Creator tier; 300,000 character limit; current-period character count 0 & Provider account check, May 2026 & Subscription/context field \\
Vercel account & Hobby plan observed & Provider account check, May 2026 & Hosting plan field \\
Supabase billing & Not recovered locally & Management API billing access returned unavailable/forbidden & Requires invoice/admin export \\
OpenAI organization costs & Not recovered locally & Costs endpoint unavailable/forbidden with available key & Requires invoice/admin export \\
\end{longtable}

Invoice reconciliation should separately report subscriptions, API invoices, provider credits, VPS cost, Vercel, Supabase, OpenRouter, and other services. The backup archive may contain credentials and private project material; it should be described in aggregate for reproducibility but not released publicly.

\section{Supplementary Figure S1. Persistent Agentic Research Environment Architecture}

This conceptual figure maps the human researcher, agent runtime, memory layer, tools, external APIs, scheduled routines, specialized agents, artifact surfaces, and governance layer. It is retained in the supplement because it explains system structure but is not itself an outcome.

\begin{figure}
\centering
\includegraphics{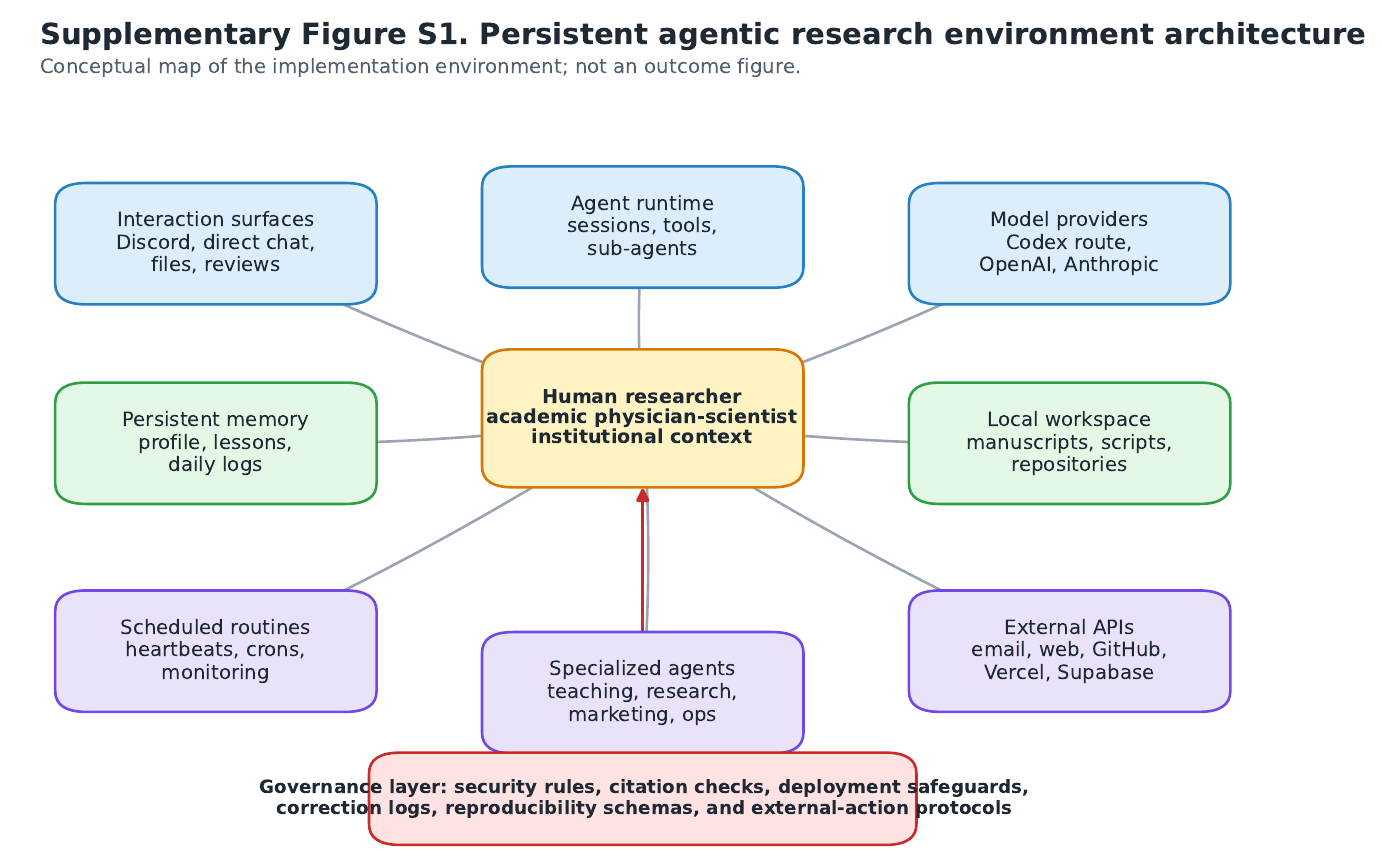}
\caption{Supplementary Figure S1. Persistent agentic research environment architecture}
\end{figure}

Source files: \texttt{figures/supplementary-figure-s1-persistent-agentic-architecture.svg}, \texttt{figures/supplementary-figure-s1-persistent-agentic-architecture.pdf}, and \texttt{figures/supplementary-figure-s1-persistent-agentic-architecture.png}.

\section{Supplementary Figure S2. Active-Time Sensitivity}

This figure reports the empirical sensitivity analysis for active-system time estimation. Panel A plots the distribution of raw inter-event gaps, clipped at 180 minutes for display. Panel B reports active-system hours and cluster counts under 15-, 30-, 45-, 60-, and 90-minute caps.

\begin{figure}
\centering
\includegraphics{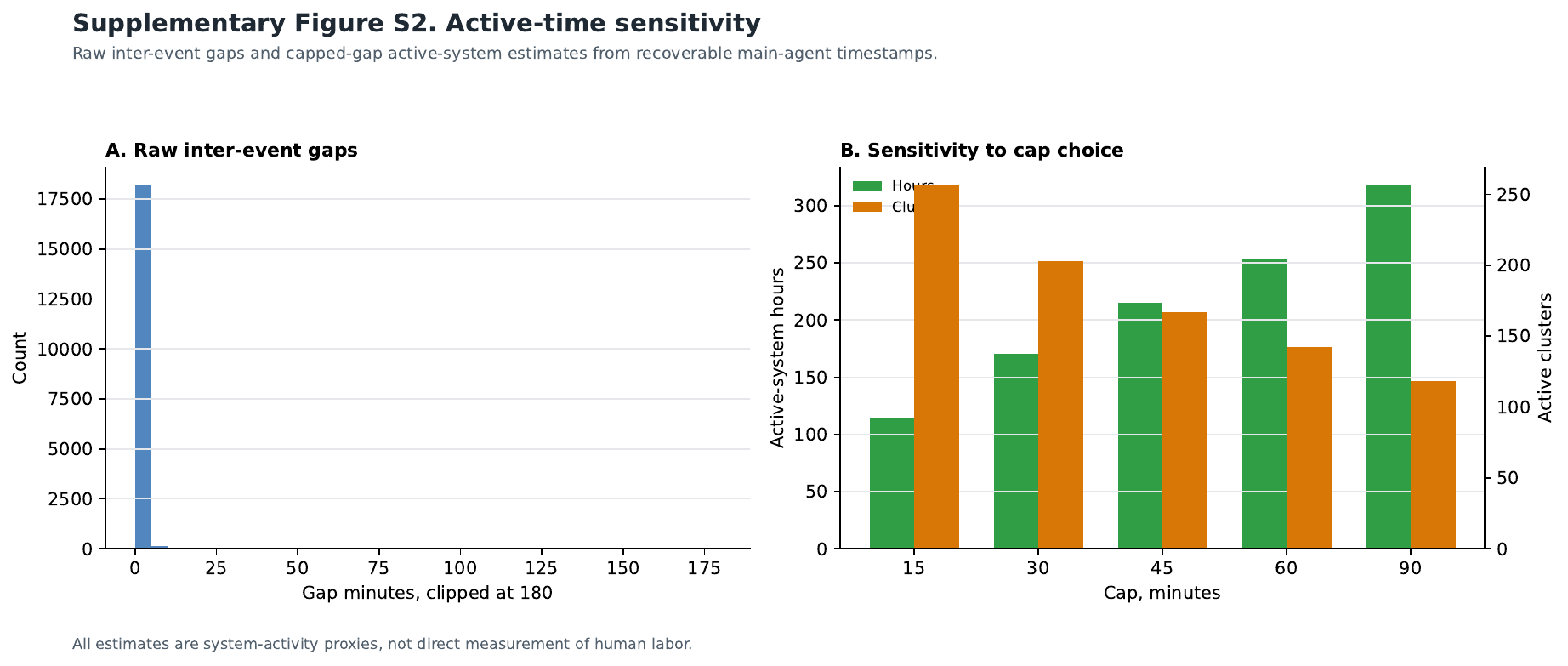}
\caption{Supplementary Figure S2. Active-time sensitivity}
\end{figure}

Source files: \texttt{figures/supplementary-figure-s2-active-time-sensitivity.svg}, \texttt{figures/supplementary-figure-s2-active-time-sensitivity.pdf}, \texttt{figures/supplementary-figure-s2-active-time-sensitivity.png}, and \texttt{figures/supplementary-figure-s2-active-time-sensitivity.csv}.

All estimates are system-activity proxies, not human labor measurement.

\end{document}